\documentclass[manuscript]{emulateapj}
\usepackage[footnotes]{trackchanges}
\usepackage{natbib}
\usepackage{times}

\begin{document}

\title{An unexpected detection of bifurcated blue straggler sequences in the young globular cluster NGC 2173}

\author{Chengyuan~Li\altaffilmark{1,2}, Licai~Deng\altaffilmark{3,2},
  Richard~de~Grijs\altaffilmark{4,5,2},
  Dengkai~Jiang\altaffilmark{6,7,8}, and Yu~Xin\altaffilmark{3}}

\altaffiltext{1}{Department of Physics and Astronomy, Macquarie
  University, Balaclava Road, North Ryde, NSW 2109, Australia}

\altaffiltext{2}{Department of Astronomy, China West Normal
  University, Nanchong 637002, China} 

\altaffiltext{3}{School of Astronomy and Space Science, University of
  the Chinese Academy of Sciences, 20A Datun Road, Beijing 100012,
  China}

\altaffiltext{4}{Kavli Institute for Astronomy \& Astrophysics and
  Department of Astronomy, Peking University, Yi He Yuan Lu 5, Beijing
  100871, China}

\altaffiltext{5}{International Space Science Institute--Beijing, 1
  Nanertiao, Zhongguancun, Beijing 100190, China}

\altaffiltext{6}{Yunnan Observatories, Chinese Academy of Sciences,
  Kunming 650216, China}

\altaffiltext{7}{Key Laboratory for the Structure and Evolution of
  Celestial Objects, Chinese Academy of Sciences, Kunming 650011,
  China}

\altaffiltext{8}{Center for Astronomical Mega-Science, Chinese Academy
  of Sciences, Beijing 100012, China}

\altaffiltext{*}{C. Li, L. Deng, and R. de Grijs jointly designed this
  project.}

\begin{abstract}
Bifurcated patterns of blue straggler stars in their
  color--magnitude diagrams have atracted significant attention. This
  type of special (but rare) pattern of two distinct blue straggler
  sequences is commonly interpreted as evidence of cluster
  core-collapse-driven stellar collisions as an efficient formation
  mechanism. Here, we report the detection of a
  bifurcated blue straggler distribution in a young Large Magellanic
Cloud cluster, NGC 2173.  Because of the cluster's low central stellar
number density and its young age, dynamical analysis shows that
stellar collisions alone cannot explain the observed blue straggler
stars. Therefore, binary evolution is instead the most viable
explanation of the origin of these blue straggler stars. However, the
reason why binary evolution would render the color--magnitude
distribution of blue straggler stars bifurcated remains unclear.
\end{abstract}


\section{Introduction}

Blue straggler stars (BSSs) are commonly found in old globular
clusters (GCs), which are generally characterized by ages in excess of
$\sim$10 Gyr, and in Galactic open clusters (OCs) of various ages
\cite[e.g.,][]{Ferr03a,Xin05a,Math09a,Li13a,Bald16a}. They appear to
be main-sequence (MS) stars that are significantly more massive than
the cluster's bulk population \citep{Sand53a,Stry93a}. In star
clusters, BSSs usually occupy a region that is bluer and
  brighter than the MS turnoff (MSTO) in the Hertzsprung--Russell
diagram or its observational equivalent, the color--magnitude diagram
(CMD). They tend to have an upper limit $\sim$2.5 magnitudes brighter
than the MSTO. Blue straggler stars are thought to be produced either
by direct stellar collisions \citep{Hill76a} or through the evolution
of binary systems, e.g., through mass transfer or stellar mergers
 \citep{McCr64a,Andr06a}. Direct stellar
  collisions can also involve binary--binary or binary--single star
  interactions \citep{Freg04a}, or interactions of triple stars
  through the Kozai mechanism \citep[e.g.,][]{Pere09a}. We encourage
  readers to explore \cite{Boff15a} for more details regarding our
  current understanding of BSSs.

The relative importance of direct stellar collisions versus binary
evolution in old GCs remains an open question. Nevertheless, it has
been suggested that the dominant BSS formation channel may be through
binary evolution, irrespective of the host cluster's dynamical state
\citep{Knig09a}. Without going through some special dynamical
processes, BSSs formed through any of the possible formation channels
would be expected to exhibit a featureless distribution beyond the
MSTO of their host cluster.  However, observations of two
distinct BSS populations featuring similar numbers of stars in the
CMDs of a number of old Galactic GCs offer evidence in support of the
notion that stellar collisions could be of comparable importance
\citep{Ferr09a,Dale13a,Simu14a}. Numerical simulations have shown that
zero-age collisional BSSs would lie on a locus that is clearly bluer
than that defined by the BSSs formed through binary mass transfer,
thus implying that the blue-sequence stars are likely products of
stellar collisions \citep[e.g.,][]{Ferr09a}. However, because the
formation rate of collisional BSSs depends on the local stellar number
density, only clusters that are sufficiently old and dense could
produce sufficient numbers of collisional BSSs along a cluster's MS
extension \citep{Davi04a}. Indeed, so far three GCs have been detected
that show a bifurcation in their BSS populations, two of which exhibit
cusps in their radial number-density profiles \citep{Ferr09a,Dale13a},
a feature expected to result from core collapse
\citep{Cohn80a,Lynd80a}. The only exception is the GC NGC
1261. Although it does not show evidence of a cusp in its radial
number-density profile, it has nevertheless been claimed to be in a
post-core-collapse state \citep{Simu14a}. Therefore, all of these
features observed in these old GCs may imply that the frequent stellar
collisions that produce the blue-branch stars were driven by
core-collapse events.

In this article, we report the unexpected detection of a bifurcated
BSS distribution in the Large Magellanic Cloud (LMC) cluster NGC 2173,
which is much younger and which has a much lower central stellar mass
density than the old GCs exhibiting bifurcated BSS populations. This
is the first detection of a bifurcated distribution of BSSs in a
cluster that is much younger than the old GCs. Dynamical calculations
carried out for this cluster show that stellar collisions are unlikely
responsible for the observed BSSs; therefore, binary evolution may
indeed be the only viable scenario. However, why BSSs formed through
binary evolution would form a bifurcated distribution in a cluster's
CMD remains an open question.

This article is organized as follows. Section \ref{S2} includes the
details of the observations and the data reduction. In Section
\ref{S3} we present our main results, which we briefly discuss in
Section \ref{S4}. Section \ref{S5} provides a summary of the main
results and our conclusions.

\section{Data Reduction}\label{S2}
\subsection{Photometry and Data Processing}

The cluster NGC 2173 was observed with the {\sl Hubble Space
  Telescope} ({\sl HST}) using the Wide Field Camera 3/Ultraviolet and
Visual Channel (WFC3/UVIS) as part of program GO-12257 (PI:
L. Girardi). The observations of NGC 2173 were obtained through the
F336W and F814W filters, with total exposure times of 1980 s and 1520
s, respectively. For the observations in both filters, we used the
WFC3 module of the {\sc dolphot2.0}
package\footnote{http://americano.dolphinsim.com/dolphot/} to perform
point-spread-function (PSF) photometry on the flat-fielded data frames
(referred to as `\_flt'). {\sc dolphot2.0} automatically calculates a
sky map and combines observations with short and long exposure times
into a final stellar catalog.

We processed the raw stellar catalog as follows. We first selected all
detected objects flagged as `good stars' by {\sc dolphot2.0}. Next, we
adopted a filter employing a sharpness constraint based on the
sharpness parameter calculated by {\sc dolphot2.0}, which enabled us
to remove unusually concentrated objects (such as cosmic rays) or
extended sources (such as background galaxies). A perfect star should
have a sharpness equal to zero. We only selected objects with $-0.2<$
sharpness $<0.2$ in both frames. The `crowding' parameter quantifies
how much brighter an object would have been had nearby stars not been
fitted simultaneously (it is expressed in units of magnitudes). We
further only selected objects with crowding $\leq 0.5$ mag in both
frames. Because the positions of the BSSs in the CMD are very
different from those expected for cosmic rays or extended sources,
while they are also very bright,  we confirmed that our
  data reduction process would not remove any of detected BSS
  candidates.

\begin{figure*}[htbp!]
\begin{center}
\includegraphics[width=18cm]{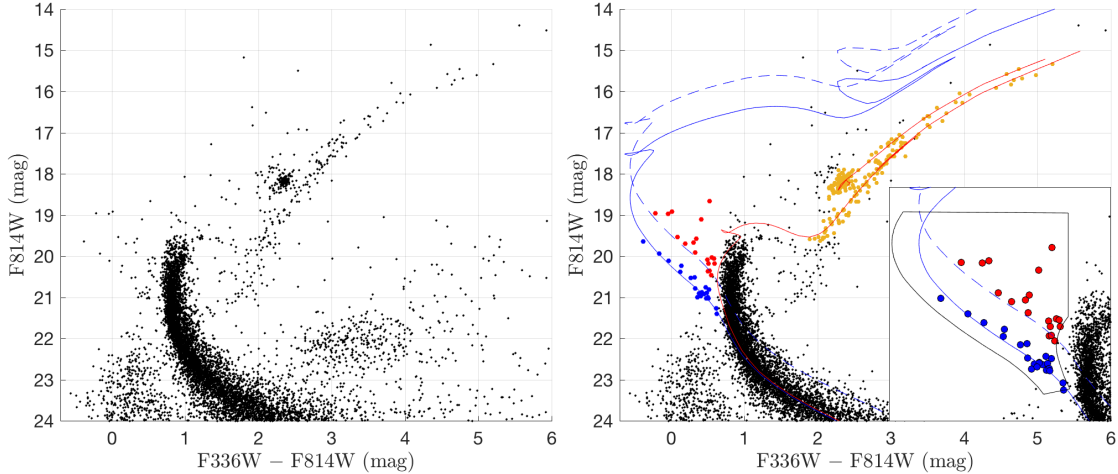}
\caption{Unprocessed (left) and final (right) CMDs of NGC 2173. Blue,
  red circles: blue, red-branch BSSs; orange circles: selected RGB,
  RC, and AGB stars. The red and blue solid lines are best-fitting
  isochrones with ages of $\log{(t \mbox{ yr}^{-1})}=9.20$ and 8.40,
  respectively. The blue dashed line is the locus of the equal-mass
  binary sequence corresponding to the young isochrone. The inset in
  the right-hand panel highlights the region used for selecting
  BSSs. {\bf (Stellar catalogs behind this figure are online accessible.)}}\label{F1}
\end{center}
\end{figure*}

We estimated the differential reddening suffered by each star by
application of the method proposed by \cite{Milo12a}. We found that
the reddening variations across the full observational field of NGC
2173 range from $E(B-V)=-0.03$ mag to 0.09 mag. The resulting
reddening-corrected CMD does not show any significant differences
compared with the original CMD. In Fig. \ref{F1} we present the raw
CMD (left panel), as well as the processed CMD of the entire image of
NGC 2173. The bifurcated BSS populations are obvious in both CMDs.

\subsection{Isochrone Fitting}

We used the PARSEC isochrones \citep{Bres12a} to fit our
observations. We used two isochrones to fit the bulk of the cluster
population and the BSSs, respectively. For the bulk stars, we adopted
an age of $\log{(t \mbox{ yr}^{-1})}= 9.20$ (1.58 Gyr), a metallicity
of $Z= 0.008$, an extinction of $A_{V}= 0.155$ mag, and a distance
modulus of $(m-M)_0= 18.40$ mag.  Because of the small
  number of BSSs and the lack of an apparent MSTO region, adopting a
  precise age for the BSSs is difficult. We forced the isochrone to
  fit the brightest blue-branch BSS. This leads to an adopted
  (maximum) age of $\log{(t \mbox{ yr}^{-1})}= 8.40$ (250
  Myr). Fig. \ref{F1} shows that the brightest blue-branch BSS is very
  close to the zero-age MS (ZAMS), which means that any isochrone with
  an age younger than 250 Myr would be able to adequately represent
  these blue-branch stars. The adopted age of 250 Myr is therefore
  simply an upper limit. Our adopted best-fitting parameters
 for the bulk of the cluster's stellar population are
close or identical to those of \cite{Groc07a} and \cite{Milo09a}. The
best-fitting isochrones are shown in the right-hand panel of
Fig. \ref{F1}. The red-branch BSSs are characterized by a lower
boundary which is close to the locus of the equal-mass binary
sequence. One can easily separate the two BSS branches through visual
inspection, as shown in the right-hand panel of Fig. \ref{F1}
(insets).

\subsection{Determinations of the Different Stellar Samples}

We selected BSSs as follows: (a) their magnitudes cover the range
$18.0\leq m_{\rm F814W} \leq 21.5$ mag; (b) they are brighter than the
250 Myr-old isochrone plus 0.1 mag in the F814W filter (on average,
this is the 3$\sigma$ level defined by the photometric uncertainties);
(c) they are bluer than the 1.58 Gyr-old isochrone minus 0.15 mag in
($m_{\rm F336W} - m_{\rm F814W}$), corresponding to, on average, the
3$\sigma$ color uncertainty level. The total number of
  selected BSSs is 40.

As shown in Fig. \ref{F1}, the BSSs in NGC 2173 exhibit a clear
bifurcation in the cluster's CMD. For each BSS, we calculated its
relative distance to the 250 Myr-old isochrone, as well as to the
corresponding equal-mass MS--MS binary locus. BSSs which were located
closer to the isochrone for single stars were assigned to the blue
sequence (see the blue circles in Fig. \ref{F1}), while those closer
to the equal-mass binary locus were included in our red-sequence
sample (the blue circles in Fig. \ref{F1}).

We also selected a sample containing most red-giant branch (RGB), red
clump (RC), and asymptotic giant branch (AGB) stars. We first
identified the position of the bottom of the RGB in the CMD. Stars
brighter than this position and located in the 3$\sigma$ region
bracketing the isochrone were selected as sample stars (see the orange
circles in Fig. \ref{F1}).

\subsection{Artificial Star Tests}

Some artificial effects, like line-of-sight blending of two stars or
the diffraction spikes of bright giant stars, may mimic a broadened or
even a split BSS population. To inspect if any of our
  BSSs might have been affected by the spikes of bright stars, we
  examined their `roundness' parameter calculated by {\sc
    dolphot2.0}. A perfect star should have a roundness of zero. If an
  object is extended or has been affected by diffraction spikes, its
  roundness will be significantly larger than that for `good'
  stars. The largest roundness value for our BSSs is 0.265, which is
  smaller than the equivalent values for 80\% of all detected objects;
  39 of the 40 observed BSSs have roundness values below 0.103, which
  is better than the values for 90\% of all detected objects.
We additionally explored the importance of these effects
  globally through artificial star tests. The principle of this
method consists of generating a large number of artificial stars
characterized by the same PSF as the observational data, adding them
to the raw data image, and then using the same approach as for the
sample of real stars to recover them. By comparing their output CMD to
the corresponding input diagram, we can evaluate how artificial
effects (including photometric uncertainties) change the morphology of
the original color--magnitude distribution.

To do this, we generated two different artificial star samples
composed of artificial stars characterized by an initial
color--magnitude distribution defined by the 250 Myr and 1.58 Gyr-old
isochrones, both with a Kroupa-like mass function \citep{Krou01a}. To
optimize our calculation time, for the young artificial stellar sample
we only generated stars that were brighter than $m_{\rm F814W} \leq
23$ mag, because the bottom morphology of its MS is almost identical
to that of the old artificial stars. Our young and old artificial
stellar samples contained 70,000 and 700,000 stars,
respectively. Their spatial distributions were homogeneous. To avoid a
situation in which the artificial stars dominate the background and
crowding levels, we only added 100 artificial stars to the raw image
at any one time. This means that for the young and old artificial
stellar samples, we repeated this procedure 700 and 7000 times,
respectively.

\begin{figure}[htbp!]
\begin{center}
\includegraphics[width=9cm]{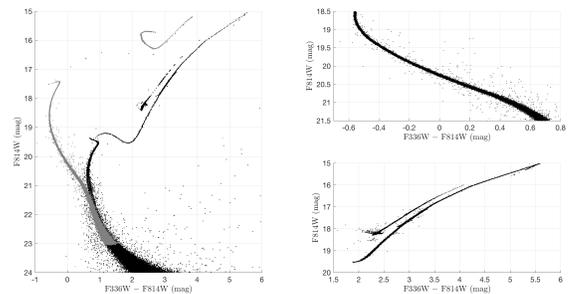}
\caption{(Left) Output CMDs of the young (grey) and old (black)
  artificial stellar samples. (Top right) Output CMD of the young
  artificial stars with the same magnitude range as the observed
  BSSs. (Bottom right) Output CMD of the artificial RGB, RC, and AGB
  stars. To optimize image space, we only exhibit a fraction of our
  artificial stars.}\label{F2}
\end{center}
\end{figure}

We present the output CMDs of both the young and old artificial stars
in the left-hand panel of Fig. \ref{F2}. We found that the recovered
color--magnitude diagram shows a MS of roughly the same width as the
real MS, as well as a broadened region toward the red side of the
MS. The latter is caused by stellar blending. We specifically
investigated the output CMDs of both regions for BSSs and RGB, RC, and
AGB stars, as shown in the top and bottom right-hand panels of
Fig. \ref{F2}. We did not find any bifurcated patterns in these
regions. Our detection of bifurcated BSS populations is therefore
unlikely caused by artificial effects.

The artificial stellar catalog also allowed us to derive a stellar
completeness map. Artificial stars that meet the criteria below are
considered stars that would contribute to the sample's incompleteness:
\begin{enumerate}
\item{Stars that do not return any photometric result.}
\item{Stars not defined as a `good' star by {\sc Dolohot2.0}.}
\item{Stars with crowding parameter $> 0.5$ mag.}
\item{Stars with sharpness parameter $< -0.2$ or $> 0.2$.}
\end{enumerate}

The completeness levels will be used to correct the derived stellar
number-density profile, as well as any radial distributions of the
numbers for different stellar samples.

\subsection{Determination of the Cluster's Global Parameters}

\begin{figure*}[htbp!]
\begin{center}
\includegraphics[width=18cm]{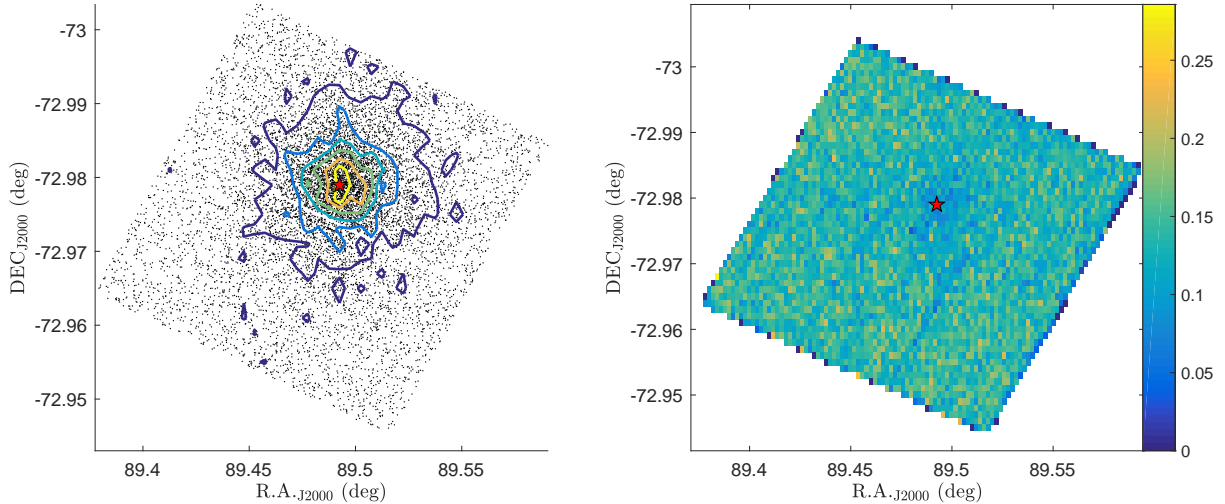}
\caption{(Left) Spatial stellar distribution in the observed field of
  NGC 2173. The red pentagram represents the cluster
  center. Number-density contours have been superimposed. (Right)
  Completeness map, with completeness levels indicated by the color
  bar, for all artificial stars.}\label{F3}
\end{center}
\end{figure*}

To determine the coordinates of the cluster center, we generated
number-density contours for the cluster and determined the coordinates
where the local number densities reached the largest values. We
adopted the latter as the cluster's center. The resulting center
coordinates are $\alpha_{\rm J2000}=-5^{\rm h}57^{\rm m}58.32^{\rm s}$
and $\delta_{\rm J2000}=-72^{\circ}58'44.40''$. The left-hand panel of
Fig. \ref{F3} presents the spatial stellar distribution, the
number-density contours, and the resulting cluster center. We realize
that blending affects the number-density level at different positions,
especially in the cluster's central region. In the right-hand panel of
Fig. \ref{F3}, we compare the position derived for the cluster center
with the completeness map. The cluster center is indeed located in a
region of relatively low stellar completeness, indicating that our
derived center position is reliable. In addition, our center
coordinates are very close to those adopted by \cite{Baum13a}.

The stellar completeness map shown in the right-hand panel of
Fig. \ref{F3} shows a very low level of completeness ($<30$\%). This
is expected, because we have generated an artificial stellar sample
that contains a large number of low-mass stars (based on the Kroupa
mass function adopted). Almost all of these low-mass (and thus faint)
stars cannot be detected. If we were to constrain the stellar sample
to the range $m_{\rm F814W}\leq 24$ mag, the overall completeness
would be higher than 90\%.

After obtaining the cluster center and sample completeness levels, we
calculated the stellar number-density profile using all detected stars
with $m_{\rm F814W} \leq 24$ mag. We first used the cluster center to
define annular rings with intervals of 1 pc\footnote{At the distance
  of the LMC, 1 arcsec is roughly equal to 0.24 pc.}. The stellar
number density in each ring is the ratio of the observed number of
stars in the ring and the ring's area, corrected for the level of
incompleteness,
\begin{equation}
\rho(r)=\frac{N(r)}{f(r)A(r)},
\end{equation}
where $N(r)$ is the number of stars in the ring at radius $r$, $f(r)$
is the corresponding completeness in the ring, and $A(r)$ is the
ring's area.

\begin{figure}[htbp!]
\begin{center}
\includegraphics[width=9cm]{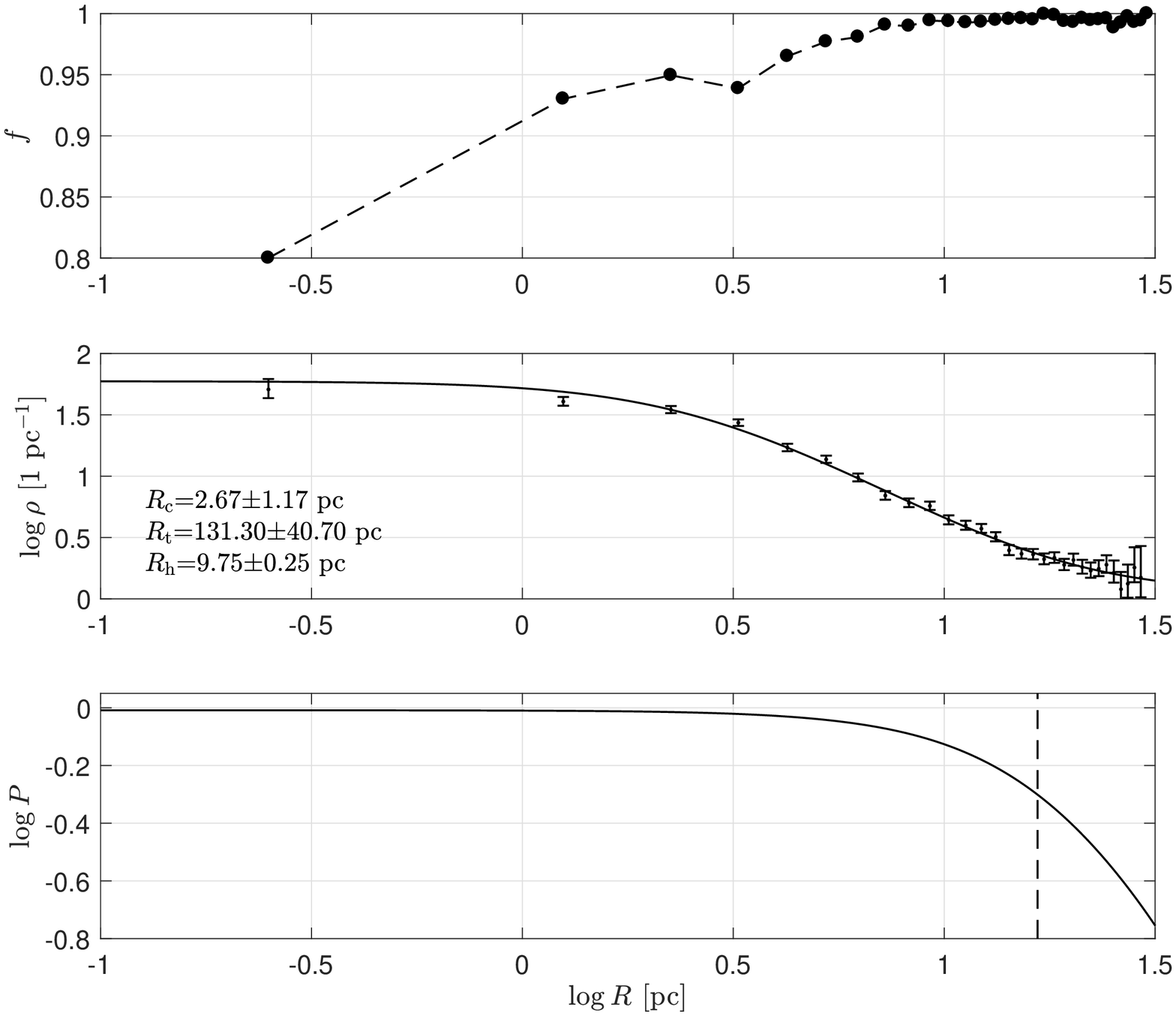}
\caption{(Top) Stellar completeness as a function of radius. (Middle)
  Stellar number-density profile and best-fitting King model. (Bottom)
  Estimated cluster membership probability as a function of
  radius.}\label{F4}
\end{center}
\end{figure}

The stellar completeness as a function of radius, as well as the
resulting number-density profile, are shown in Fig. \ref{F4}. We next
used the empirical King model \citep{King62a} and a constant
background field population, $b$, to fit their profiles,
\begin{equation}
\rho(r)=k\left[\frac{1}{\sqrt{1+(r/r_{\rm
        c})^2}}-\frac{1}{\sqrt{1+(r_{\rm t}/r_{\rm c})^2}}\right]+{b},
\end{equation}
where $r_{\rm c}$ and $r_{\rm t}$ are the core- and tidal radii,
respectively. If we assume that the background stars are homogeneously
distributed across the entire image, then $b$ is a constant which
represents the number density of the background; $k$ is a
normalization coefficient. Our best-fitting core- and tidal radii are
$r_{\rm c}=2.67$ pc and $r_{\rm t}=131.30$ pc, respectively. Because
of the small {\sl HST}/WFC3 field, the full image actually only
includes about a quarter of the cluster's tidal radius; the maximum
cluster radius covered is only $\sim$32 pc. Based on the best-fitting
King model, we also derived the radius that contains half the number
of stars, i.e., $r_{\rm h}=9.75$ pc. Our result is consistent with
that derived by \cite{Mcla05a}.

The commonly adopted approach to subtract the field stellar
contribution involves selecting a nearby region as reference field,
followed by randomly removing a similar color--magnitude distribution
from the observed cluster+field CMD as that covered by the field stars
only. However, the large size of NGC 2173 and the small size of the
observed field cause problems in this regard. We cannot properly
account for the effects of field-star contamination based on the
cluster's CMD. Indeed, as we already showed in Fig. \ref{F1}, the
bifurcated pattern of BSSs is already obvious in the raw CMD of the
entire observed region. Field stars of different ages and
metallicities are very unlikely to give rise to such a feature,
however. In this article, we did not subtract any `field' stars from
the observed CMD. Instead, we statistically evaluated the
contributions of field stars at different radii. We defined the
cluster member probability of each star as the ratio of the local
densities of the star cluster and the background field,
\begin{equation}
P(r)=\frac{\rho(r)-b}{\rho(r)}.
\end{equation}
The profile of the cluster membership probability is presented in the
bottom panel of Fig. \ref{F4}. As expected, a large fraction of the
image is actually dominated by cluster stars. We found that for radii
$r\leq16.65$ pc, the observed numbers of stars are dominated by the
cluster ($P\geq50$\%). Below, we will specifically limit our study to
BSSs found within this radius.

\label{sec:real}

\section{Main Results}\label{S3}
\subsection{Significance}

As shown in the right-hand panel of Fig. \ref{F1}, the bifurcated BSS
pattern in the CMD of the NGC 2173 field is indeed obvious. The
numbers of the red- and blue-branch stars are comparable, with 18 of
them belonging to the red branch and 22 associated with the blue
branch.  We first evaluate the tightness of our blue- and red-branch BSSs by 
comparing their color-magnitude distributions to the best fitting isochrone (or the 
equal-mass binary sequence for red-branch BSSs). We calculate the 
color deviation for each blue- and red-branch stars to the best fitting isochrone 
(equal-mass binary sequence). We found the distribution of color deviations for 
blue-branch BSSs can be well described by a Gaussion function, with a standard 
deviation of $\Delta_{\rm c}$=0.060$\pm$0.033 mag, which is consistent with the typical 
photometric error for BSSs $\Delta_{\rm c}$=0.041$\pm$0.002 mag, the latter is 
determined from the CMD of artificial stars. This means the blue-branch BSSs 
can be well explained by a population of single stars with photometric errors. 
In contrast, the distribution of color deviations for red-branch BSSs cannot be described 
by a Gaussion distribution. To further quantify the dispersion of red-branch 
stars, we calculate the root-mean-square (RMS) of their color deviations, 
we found the RMS of red-branch BSSs is more than three times that of 
blue-branch BSSs (0.23 mag vs. 0.07 mag). In summary, the blue-branch 
BSSs are well associated with the best fitting isochrone. The red-branch 
BSSs are more dispersed than the blue-branch BSSs.
  
\begin{figure*}[htbp!]
\begin{center}
\includegraphics[width=18cm]{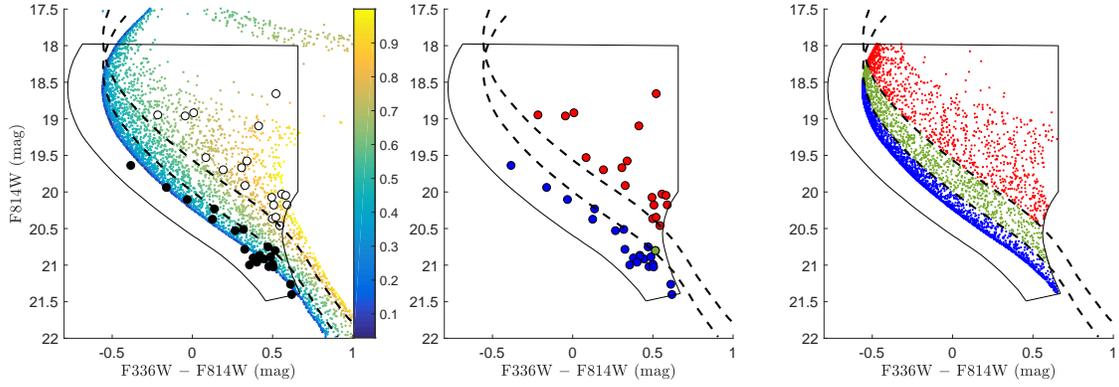}
\caption{(Left) CMDs of BSSs combined with our synthetic stellar
  populations. The filled and open circles represent the blue- and
  red-branch BSSs, respectively. The mass ratios of our synthetic
  stars (binaries) are indicated by the color scale. The black dashed
  curves were adopted to roughly define the gap region between the two
  branches of BSSs. (Middle) CMDs of BSSs located in the blue branch
  (blue circles), the red branch (red circles), and the gap (green
  circle). (Right) As the middle panel, but for the synthetic
  stars.} \label{F5}
\end{center}
\end{figure*}

We then investigate the significance of the bifurcated pattern of BSSs, to do this we 
generated a synthetic stellar sample with an age of 250 Myr located in the BSS region. 
For each of them, we randomly assigned a binary component associated with the 
1.58 Gyr-old stellar population, with the relevant mass ratio selected randomly from zero 
to unity. All these binary systems are unresolved in our simulation. The synthetic CMD is shown 
in the left-hand panel of Fig. \ref{F5}, where the mass ratios are indicated by the 
color bar. 

Intriguingly enough, unlike the binary sequence below the MSTO region, 
the high mass-ratio binary sequence in the BSS region is not parallel to the ZAMS. 
The binaries are significantly dispersed toward the red, which is similar to the observed 
red-branch BSSs. The reason for unparalleled high mass-ratio binary sequences is 
because the secondary stars associated with these high mass-ratio binary systems have already
  evolved off the MS (or may be ready to leave the MS), therefore these binary systems are 
  no longer MS-MS binaries.

We divided both the observed and synthetic samples into three parts (see the 
black dashed lines in Fig. \ref{F5}) which roughly define the regions for the blue- and 
red-branch stars, and the middle gap. In our adoption, only one BSS is
  located between these two lines. We then randomly selected 40 synthetic stars 
  from our artificial stars. We repeated this procedure 10,000
  times and counted how many times we were left with no or only one
  star in the gap. We found that only five times did we obtain a gap
  in our distribution of 40 synthetic stars. This means that the
  probability that the detected bifurcated pattern may be a stochastic
  feature is only 0.05\%,  corresponding to a significance level of greater 
  than 3$\sigma$. We also constrained the synthetic stellar sample with 
  a uniform color-magnitude distribution in our BSS region. In that case, the probability 
of the observed feature representing the result of stochastic sampling became 0.95\%
  (95/10,000 times),  which equals to a significance level of 2--3$\sigma$.

\label{sec:real}

\subsection{Radial BSS Distribution}

\begin{figure*}[htbp!]
\begin{center}
\includegraphics[width=18cm]{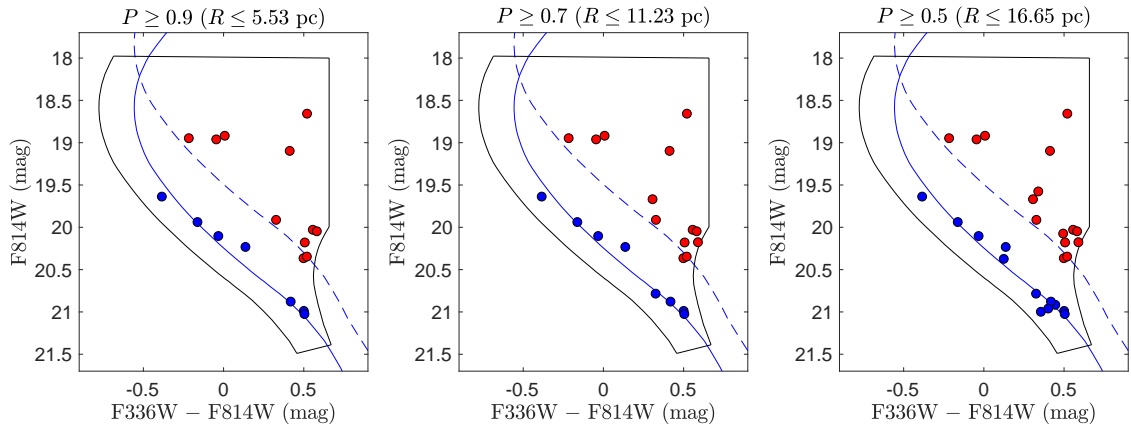}
\caption{CMDs of BSSs with different cluster membership
  probabilities. From left to right, $P\geq90$\%, 70\%, and
  50\%.}\label{F6}
\end{center}
\end{figure*}

If we focus on the region characterized by stellar cluster membership
probabilities greater than 50\% ($r<16.65$ pc), the numbers of BSSs
associated with the red and blue branches are 12 and 15,
respectively. (Note that this relates to the number of BSSs in the
entire observed field.) If we further constrain the sample of BSSs to
cluster membership probabilities exceeding 90\% ($r<5.53$ pc), the
bifurcated feature is still significant, and the numbers of red- and
blue-branch stars are 11 and 7, respectively. Fig. \ref{F6} presents
three CMDs for BSSs with cluster membership probabilities $P\geq90$\%,
$P\geq70$\%, and $P\geq50$\%.

\label{sec:real}

\begin{figure*}[htbp!]
\begin{center}
\includegraphics[width=18cm]{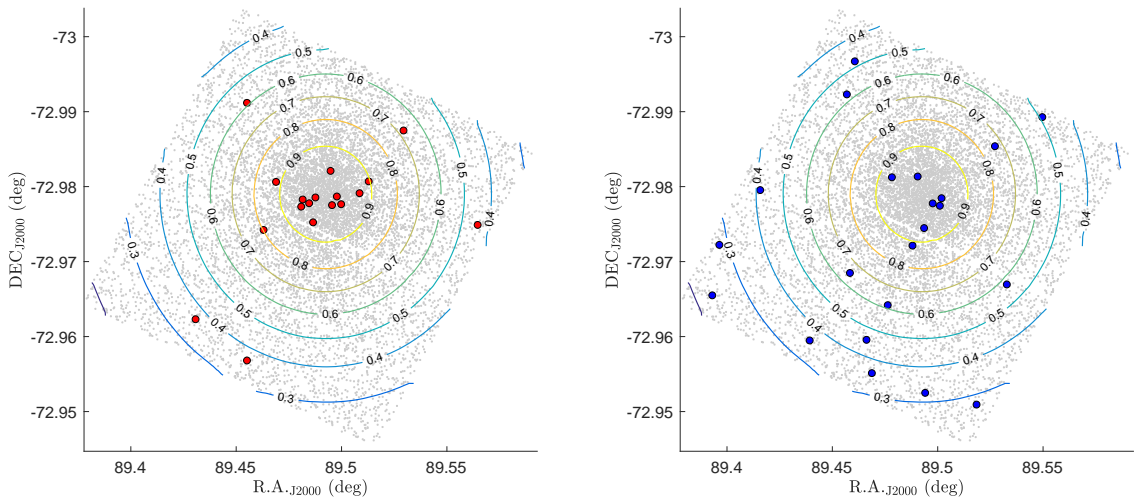}
\caption{Spatial distributions of (left) the red- and (right)
  blue-branch BSSs. Radii characterized by different cluster
  membership probabilities are indicated.}\label{F7}
\end{center}
\end{figure*}

In the spatial distributions of the blue- and red-branch BSSs (Fig.\ref{F7}), the 
red-branch BSSs are likely more centrally concentrated than the blue-branch 
counterparts. Although the majority of blue-branch stars are located inside 
the radius characterized by $P\geq50$\% (12/22), a significant fraction of
blue-branch stars (10/22) is located beyond the cluster-dominated
region. As a comparison, more than 13 of 18 red-branch stars are located 
inside the $P=50\%$ radius, only 5 of them are located beyond this radius. 
However, because the number of our BSSs is very small, it is hard to 
conclude that the red-branch stars are obviously more segregated than 
blue-branch stars. Our statistical test report that the chance for these 
two branch stars were drawn from the same parent spatial distribution 
is 5.35\%. That means the significance for their difference in central 
concentration is 94.65\%, corresponding to a 1--2$\sigma$ level significance.

It is possible that most of these latter stars (for both the
red- and the blue-branch BSSs) are actually field stars. However,
because all observed BSS stars are distributed into two distinct
sequences (see Fig. \ref{F1}), it would be puzzling if field stars of
different ages and metallicities would form a clearly bifurcated
pattern in the CMD. Another explanation is that most of these stars
are genuine BSSs formed in the cluster environment, while the sample
stars at larger radii were formed in the core and subsequently ejected
to the outer regions by the recoil of stellar interactions
\citep{Sigu94a}. These stars, simply based on their spatial 
distribution, looks like field stars, but are actually cluster members.
Alternatively, they may be BSSs that are not yet
fully segregated \citep{Ferr12a}. Since we lack kinematic information
for these objects (such as proper motions or radial
  velocities), we cannot draw any definite conclusions as regards the
origin of these BSSs.

\begin{figure}[htbp!]
\begin{center}
\includegraphics[width=9cm]{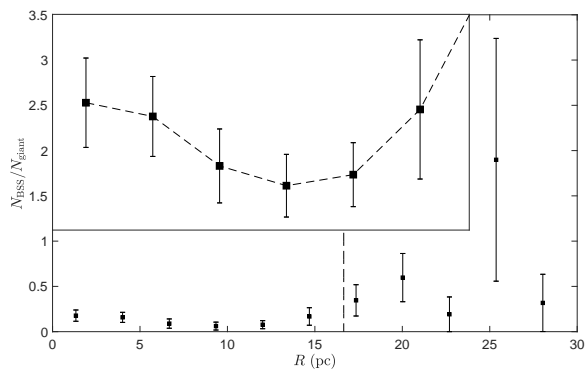}
\caption{BSS-fraction profile (normalized to the number of giant
  stars), corrected for stellar incompleteness. The black dashed line
  indicates the radius where $P = 50$\%. The number-fraction profile
  within this radius is shown in the inset.}\label{F8}
\end{center}
\end{figure}

We explored the radial behavior of these BSSs relative to the bulk
population of giant stars (RGB, RC, and AGB stars). We first took both
the red- and blue-branch BSSs as our whole sample. We divided the
selected sample of BSSs and giant stars into eleven radial bins with
intervals of 2.67 pc (i.e., equal to the cluster's core radius). For
each radius, we calculated the number ratio of the BSSs and the giant
stars, corrected for stellar incompleteness. Again, stellar
completeness was estimated by exploring the difference between the
output and the input artificial stellar samples, as shown in the
right-hand panels of Fig. \ref{F2}. The radial profile of the number
fraction of the BSSs and the giant stars of NGC 2173 is presented in
Fig. \ref{F8}.

As shown in Fig. \ref{F8}, the BSS number-fraction profile is noisy in
the outer regions ($r> 17$ pc), because field stars may be dominant in
this region (i.e., $P<50$\%). The BSS number-fraction profile in this
region may not provide us with any useful information. Instead, we
explored their number-fraction profile in the cluster-dominated region
($P<50$\%, $r\leq16.65$ pc). We found that the BSSs' number-fraction
profile is much smoother, exhibiting a seemly bimodal distribution as a
function of radius. The minimum number fraction is located between
three and four core radii (8.01--10.68 pc). However, we searched for 
evidence of a bimodal spatial distribution of the entire BSS sample 
\citep[e.g.,][]{Ferr12a}, and did not find one at 2$\sigma$ level of significance.

We want to emphasize that in this work we do not find a proper way 
to precisely estimate the effect of field contamination. Because we lack 
the information of kinetics for our BSSs (such as their radial velocities or 
proper motions). Because of the large distance to NGC 2173, the 
kinematics of our BSSs can be only achieved using next-generation 
telescopes. Readers should be cautious when reach any conclusions 
based on their radial distributions.

\section{Discussion}\label{S4}

We first discuss the possible origin of the BSSs in NGC 2173. As shown
in Figs \ref{F1} and \ref{F6}, all observed red-branch BSSs are
located above the equal-mass binary locus, while the blue sequence
resembles the cluster's ZAMS. As already pointed out,
  adopting a continuous mass-ratio distribution for a sample of binary
  systems will unlikely produce such a bifurcated
  pattern. Low-mass-ratio binaries should have filled the gap between
  the two BSS branches \citep[for example,
    see][]{Duqu91a,Milo12a,Li13b}. In addition, as shown in
  Fig. \ref{F1}, NGC 2173 does not show any bifurcated feature in the
  MS region, further supporting the suggestion that MS--MS binaries
  are not responsible for the observed bifurcated BSS pattern.

For these reasons, we suggest that this pattern has nothing to do with
the `single-star versus MS--MS binary relation' as commonly found for
cluster MSs. As shown in Fig. \ref{F5}, the large dispersion and the
position of these red-branch BSSs would imply that they are likely all
high mass-ratio binaries. If their companions belong to the old
stellar population, they should be more evolved than ZAMS stars, e.g.,
expanding MSTO stars, subgiant stars, or faint red-giant stars (bright
red-giant stars will dominate the total flux of such a binary system
and render its photometric locus closer to the RGB rather than the BSS
region). This seems to make sense, because a binary system containing
an evolved component is more likely to be involved in a mass-transfer
process between its member stars, thus producing a rejuvenated BSS. If
confirmed, this may imply that a large fraction of these red-branch
BSSs are W Ursa Majoris-like objects, similar to the results of
\cite{Ferr09a} for the bifurcated BSSs in the GC M30.

However, although bifurcated BSS populations have also been observed
in three old Galactic GCs, where the blue sequences were attributed to
core-collapse-driven collisions \citep{Ferr09a,Dale13a,Simu14a}, it is
unlikely that these blue-sequence BSSs result from direct stellar
collisions. Using the equation introduced by \cite{Davi04a}, the
number of expected collisional BSSs that would be produced in a
stellar system over the last 1 Gyr is,
\begin{equation}
N_{\rm col}=0.03225\frac{f^2_{\rm mms}N_{\rm c}n_{\rm c,5}r_{\rm
    col}m_{\rm BSS}}{V_{\rm rel}},
\end{equation}
where $f_{\rm mms}$ is the fraction of massive MS stars in the region
of interest. Here, `massive' is defined as a mass that is sufficient
to form a BSS when two MS stars collide{\color{red} :} $f_{\rm mms}=
0.25$ is commonly adopted \citep{Davi04a}, and $N_{\rm c} =
1.14\times10^4$ is the number of stars contained in the core
region. This latter number was evaluated as follows. We calculated the
number of stars with masses between $\sim 1.1 M_{\odot}$ and $\sim 1.5
M_{\odot}$ (corrected for stellar incompleteness). We assumed that
these stars follow a Kroupa mass function \citep{Krou01a} and then
evaluated the total number of stars by extrapolating this mass
function down to $0.08 M_{\odot}$. $n_{c,5}$ is the core density of
stars in units of $10^5$ pc$^{-3}$, which is the ratio of the number
of stars contained in the core region and the core volume, yielding
$n_{c,5}\sim0.0014$. $r_{\rm col}$ is the minimum separation of two
colliding stars in units of the solar radius, $R_{\odot}$. The average
mass of BSSs is $m_{\rm BSS}\sim2 M_{\odot}$. A MS star with this mass
would have a radius of $r\sim 1.5 R_{\odot}$ \citep{Demi91a}. We
simply adopted $3R_{\odot}$ for $r_{\rm col}$ (about twice the stellar
radius). $V_{\rm rel}$ is the relative incoming velocity of binaries
at infinity, which can be written as,
\begin{equation}
V_{\rm rel}=\sqrt{2}\sigma=\sqrt{\frac{4GM_{\rm c}}{r_{\rm c}}},
\end{equation}
where $\sigma$ is the stellar velocity dispersion in the core
region. $M_{\rm c}$ is the stellar mass of the cluster core, for which
we derived $M_{\rm c} \sim 3700$ $M_{\odot}$. The resulting central
stellar velocity dispersion for NGC 2173 is $\sigma\sim 3.5$ km
s$^{-1}$. Our derived stellar velocity dispersion for NGC 2173 is
almost twice that found by \cite{Mcla05a}. This is most likely because
we have overestimated the number of stars in the cluster's core
region; numerous low-mass stars must have evaporated from the cluster
core owing to dynamical mass segregation \citep[e.g.,][]{grijs02a}.

We finally obtained $N_{\rm col}=0.04$ for NGC 2173. This means that
only $\sim 0.01$ collisional BSSs are expected to have formed over the
past 250 Myr and were detected by us in their MS phase. We realize
that some higher-order stellar interactions, such as
binary--single-star and binary--binary interactions, may increase the
production rate of collisional BSSs. The realistic number of
collisional BSSs produced should still be of the same order, however
\citep{Davi04a}. The negligibly small number of expected collisional
BSSs formed within the last 250 Myr excludes stellar
collisions as a possible origin of the observed tight blue-branch BSS
stars. Indeed, as shown in Fig. \ref{F4}, NGC 2173 exhibits a very
extended core region. This is in contrast to the structures of GCs
exhibiting distinct BSS sequences, which often show cusps in their
central regions, apparently driven by core-collapse events
\citep{Ferr09a,Dale13a}.

Another possible explanation is that NGC 2173 may already have
experienced a core-collapse event and has remained in the
post-core-collapse bounce state, similar to that found for the old GC
NGC 1261 \citep{Simu14a}. However, it is unclear if a core-collapse
event could occur in such a young cluster. \cite{Tren10a} calculated
that the minimum core-collapse timescale for a star cluster is about
5.8 times its half-mass relaxation timescale (their Fig. 7). Based on
\cite{Mcla05a}, the half-mass relaxation timescale pertaining to NGC
2173 is at least 2.09 Gyr. Therefore, there is no obvious evidence
that supports the notion that NGC 2173 may have experienced a
core-collapse event.

Our statistical analysis shows that the observed blue-branch BSSs 
are tightly associated with the ZAMS. This fact may indicate that most 
blue-branch BSSs actually originate from single stars. However, this 
immediately leads to another conundrum. Since only stellar collisions 
or binary mergers can result in a BSS that is not presently a member
of a binary system, the only viable explanation of these blue-branch
BSSs is that they are binary merger products. However, we have already
excluded stellar collisions as an efficient channel to
produce these BSSs. If this is correct, the tight blue sequences may
indicate that most binaries merged over very short time intervals. Why
would so many binary merger events happen over a short period? The
answer to this question remains to be resolved. Future analysis
focusing on details of the cluster's dynamics, such as its mass
segregation, binary hardening, and mass exchange, may be able to shed
light on this outstanding problem.

\section{Summary}\label{S5}

In this article, we report the detection of a bifurcated BSS pattern
in the CMD of NGC 2173. This is the first detection of such a feature
in a star cluster that is much younger than the typical GC age. Our
main results and conclusions can be summarized as follows:
\begin{itemize}
\item The overall distribution of BSSs in NGC 2173 is divided into two
  sequences with a significance of equal to or greater than 3$\sigma$ level. 
  The blue-sequence BSSs are tightly associated with a
  single-age isochrone with an age of $\sim$250 Myr. 
  The red-sequence BSSs exhibit a lower boundary that is close to the 
  equal-mass MS--MS binary locus for an age of $\sim$250 Myr.
  In the CMD, the red-branch BSSs are more dispersed than 
  blue-branch BSSs
\item The red-branch BSSs are more centrally concentrated than blue-branch 
  BSSs with a significance level of close to $\sim$2$\sigma$. 
  
\item Stellar collisions are unlikely responsible for these BSSs
  because of the cluster's low stellar central number density. In
  addition, the number-density profile of NGC 2173 exhibits an
  extended core with a size of $\sim$2.67 pc, indicating that NGC 2173
  has not experienced any core-collapse event.
\item We suggest that only binary evolution is able to explain the
  presence of these BSSs. However, why binary evolution would result
  in a BSS distribution composed of two distinct sequences in the
  cluster's CMD remains unclear.
\end{itemize}

\acknowledgements{}  We thank the anonymous referee for his/her 
valuable comments. We thank J. J. Eldridge at the University of
Auckland, New Zealand, and S. Justham at the National Astronomical
Observatories, Chinese Academy of Sciences, for their insightful
suggestions. C. Li acknowledges funding support from the Macquarie
Research Fellowship Scheme. R. de Grijs, L. Deng, Y. Xin, and D. Jiang
acknowledge research support from the National Natural Science
Foundation of China through grants 11373010, 11390374, 11521303,
11573061, 11633005, U1631102, and 11661161016. R. de Grijs is grateful
for support from the National Key Research and Development Program of
China through grant 2017YFA0402702. D. Jiang also acknowledges support
from the Natural Science Foundation of Yunnan Province through grant
2015FB190.

\bibliographystyle{apj}
\bibliography{cli}

\begin{thebibliography}{}
\expandafter\ifx\csname natexlab\endcsname\relax\def\natexlab#1{#1}\fi
\bibitem[Andronov et al.(2006)]{Andr06a} Andronov, N., Pinsonneault, M.~H., \& Terndrup, D.~M.\ 2006, \apj, 646, 1160

\bibitem[Baldwin et al.(2016)]{Bald16a} Baldwin, A.~T., Watkins, L.~L., van der Marel, R.~P., et al.\ 2016, \apj, 827, 12 

\bibitem[Baumgardt et al.(2013)]{Baum13a} Baumgardt, H., Parmentier, G., Anders, P., \& Grebel, E.~K.\ 2013, \mnras, 430, 676 

\bibitem[Boffin et al.(2015)]{Boff15a} Boffin, H.~M.~J., Carraro, G., \& Beccari, G.\ 2015, Ecology of Blue Straggler Stars,  

\bibitem[Bressan et al.(2012)]{Bres12a} Bressan, A., Marigo, P., Girardi, L., et al.\ 2012, \mnras, 427, 127

\bibitem[Cohn(1980)]{Cohn80a} Cohn, H.\ 1980, \apj, 242, 765 

\bibitem[Dalessandro et al.(2013)]{Dale13a} Dalessandro, E., Ferraro, F.~R., Massari, D., et al.\ 2013, \apj, 778, 135

\bibitem[Davies et al.(2004)]{Davi04a} Davies, M.~B., Piotto, G., \& de Angeli, F.\ 2004, \mnras, 349, 129

\bibitem[de Grijs et al.(2002)]{grijs02a} de Grijs, R., Gilmore, G.~F., Johnson, R.~A., \& Mackey, A.~D.\ 2002, \mnras, 331, 245 

\bibitem[Demircan \& Kahraman(1991)]{Demi91a} Demircan, O., \& Kahraman, G.\ 1991, \apss, 181, 313 

\bibitem[Duquennoy \& Mayor(1991)]{Duqu91a} Duquennoy, A., \& Mayor, M.\ 1991, \aap, 248, 485

\bibitem[Ferraro et al.(2003)]{Ferr03a} Ferraro, F.~R., Sills, A., Rood, R.~T., Paltrinieri, B., \& Buonanno, R.\ 2003, \apj, 588, 464

\bibitem[Ferraro et al.(2009)]{Ferr09a} Ferraro, F.~R., Beccari, G., Dalessandro, E., et al.\ 2009, \nat, 462, 1028

\bibitem[Ferraro et al.(2012)]{Ferr12a} Ferraro, F.~R., Lanzoni, B., Dalessandro, E., et al.\ 2012, \nat, 492, 393 

\bibitem[Fregeau et al.(2004)]{Freg04a} Fregeau, J.~M., Cheung, P., Portegies Zwart, S.~F., \& Rasio, F.~A.\ 2004, \mnras, 352, 1 

\bibitem[Grocholski et al.(2007)]{Groc07a} Grocholski, A.~J., Sarajedini, A., Olsen, K.~A.~G., Tiede, G.~P., \& Mancone, C.~L.\ 2007, \aj, 134, 680 

\bibitem[Hills \& Day(1976)]{Hill76a} Hills, J.~G., \& Day, C.~A.\ 1976, \aplett, 17, 87 

\bibitem[King(1962)]{King62a} King, I.\ 1962, \aj, 67, 471 

\bibitem[Knigge et al.(2009)]{Knig09a} Knigge, C., Leigh, N., \& Sills, A.\ 2009, \nat, 457, 288 

\bibitem[Kroupa(2001)]{Krou01a} Kroupa, P.\ 2001, \mnras, 322, 231 

\bibitem[Li et al.(2013)]{Li13a} Li, C., de Grijs, R., Deng, L., \& Liu, X.\ 2013, \apjl, 770, L7 

\bibitem[Li et al.(2013)]{Li13b} Li, C., de Grijs, R., \& Deng, L.\ 2013, \mnras, 436, 1497 

\bibitem[Lynden-Bell \& Eggleton(1980)]{Lynd80a} Lynden-Bell, D., \& Eggleton, P.~P.\ 1980, \mnras, 191, 483 

\bibitem[Mathieu \& Geller(2009)]{Math09a} Mathieu, R.~D., \& Geller, A.~M.\ 2009, \nat, 462, 1032 

\bibitem[McCrea(1964)]{McCr64a} McCrea, W.~H.\ 1964, \mnras, 128, 147

\bibitem[McLaughlin \& van der Marel(2005)]{Mcla05a} McLaughlin, D.~E., \& van der Marel, R.~P.\ 2005, \apjs, 161, 304 

\bibitem[Meylan(1987)]{Meyl87a} Meylan, G.\ 1987, \aap, 184, 144 

\bibitem[Milone et al.(2009)]{Milo09a} Milone, A.~P., Bedin, L.~R., Piotto, G., \& Anderson, J.\ 2009, \aap, 497, 755  

\bibitem[Milone et al.(2012)]{Milo12a} Milone, A.~P., Piotto, G., Bedin, L.~R., et al.\ 2012, \aap, 540, A16 

\bibitem[Perets \& Fabrycky(2009)]{Pere09a} Perets, H.~B., \& Fabrycky, D.~C.\ 2009, \apj, 697, 1048 

\bibitem[Sandage(1953)]{Sand53a} Sandage, A.~R.\ 1953, \aj, 58, 61 

\bibitem[Sigurdsson et al.(1994)]{Sigu94a} Sigurdsson, S., Davies, M.~B., \& Bolte, M.\ 1994, \apjl, 431, L115 

\bibitem[Simunovic et al.(2014)]{Simu14a} Simunovic, M., Puzia, T.~H., \& Sills, A.\ 2014, \apjl, 795, L10

\bibitem[Stryker(1993)]{Stry93a} Stryker, L.~L.\ 1993, \pasp, 105, 1081 

\bibitem[Trenti et al.(2010)]{Tren10a} Trenti, M., Vesperini, E., \& Pasquato, M.\ 2010, \apj, 708, 1598 

\bibitem[Xin \& Deng(2005)]{Xin05a} Xin, Y., \& Deng, L.\ 2005, \apj, 619, 824

\bibitem[Zinn \& Searle(1976)]{Zinn76a} Zinn, R., \& Searle, L.\ 1976, \apj, 209, 734 
\end{thebibliography}
\end{document}